\documentclass[prl,twocolumn,aps,amssymb]{revtex4}
\usepackage{epsfig}
\begin{document}

\title{Multistability, noise and attractor-hopping: \\ The crucial role of chaotic saddles}
\author{Suso Kraut$^{(1,2)}$ and Ulrike Feudel$^{(2)}$}
\affiliation{$^{(1)}$ Institut f\"ur Physik, Universit\"at Potsdam,
Postfach 601553,D-14415 Potsdam, Germany \\
$^{(2)}$ ICBM,
Carl von Ossietzky Universit\"at, PF 2503, 26111 Oldenburg, Germany}
\date{\today}

\begin{abstract}

We investigate the hopping dynamics between different attractors in
a multistable  system under the influence of noise.
Using symbolic dynamics we find a sudden increase of dynamical entropies,
when a system parameter is varied.
This effect is explained by a novel bifurcation involving two
chaotic saddles. We also demonstrate that the transient
lifetimes on the saddle obey a scaling law in analogy to crisis.

PACS number 05.45+b\\
\end{abstract}

\maketitle
\begin{figure}[h] 
\begin{center}
\epsfig{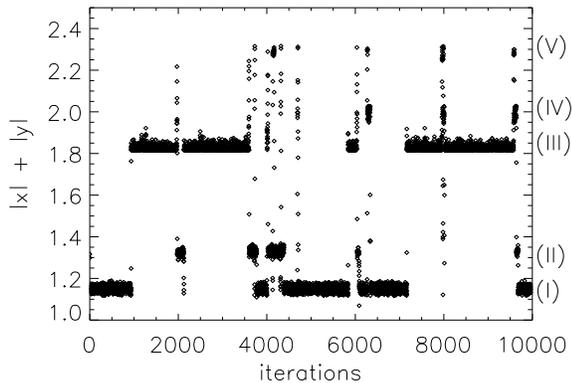}
\vspace{0.5cm}
\caption{Noisy time series of equation (\ref{coupled_log_map}) for
$\alpha = 0.73 , \gamma = 0.29$ and $\sigma = 0.012$. There are
clearly $5$  distinct almost periodic states, marked with roman
numbers on the right hand side. These states are
interrupted by bursts, where the motion takes place on a chaotic saddle.}
\label{timeseries}
\end{center}
\end{figure}
\begin{figure}[h] 
\begin{center}
\epsfig{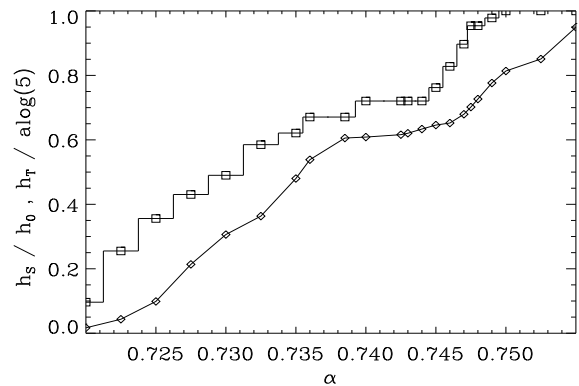}
\vspace{0.5cm}
\caption{The Shannon entropy ($\diamond$) and the topological entropy
($\Box$) of the symbol sequence  generated from equation
(\ref{coupled_log_map}) under variation of the nonlinearity parameter
$\alpha$. Both quantities are normalized to 1.}
\label{entropy}
\end{center}
\end{figure}
\begin{figure}[h]
\begin{center}
\epsfig{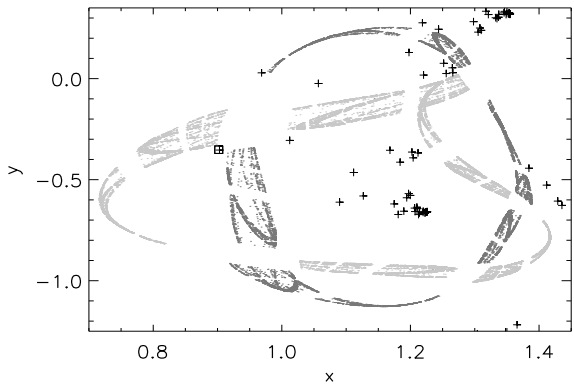}
\vspace{0.25cm}
\caption{Two chaotic saddle rings in different gray scale together
with the $4$-fold preimages $(+)$
of the repellor $R$ marked with a $\Box$ for $\alpha = 0.740$.}
\label{saddle1}
\end{center}
\end{figure}
\begin{figure}[h]
\begin{center}
\epsfig{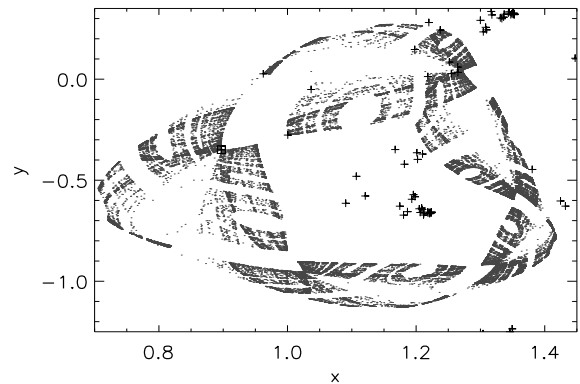}
\vspace{0.25cm}
\caption{Chaotic saddle together with the $4$-fold preimages $(+)$
of the repellor $R$ marked with a $\Box$ for $\alpha = 0.7430$.}
\label{saddle2}
\end{center}
\end{figure}

Systems with a large number of coexisting stable states have been the
subject of increasing interest recently.
This multistable behavior occurs in many different fields
like optics \cite{Wiesenfeld:1990}, chemistry
\cite{Marmillot:1991}, neuroscience \cite{schiff:1994},
semiconductor physics \cite{Prengel:1994}, plasma physics
\cite{Chern:1991} and coupled oscillators \cite{kim:1997b}. 
Moreover, addition of noise to multistable systems has led to several
interesting phenomena, like noise-induced preference of attractors 
\cite{Kaneko:1997,Kraut:1999}, directed diffusion \cite{hondou:1995},
noise-enhanced multistability \cite{kim:1997a} and chaotic itinerancy
\cite{ikeda:1989}. The latter effect, which consists of hopping
between different attractors caused by the noise, has also been
observed experimentally in an optical system \cite{Arecchi:1990}.
\\
In this Letter we focus on the dynamics of the  attractor-hopping
process by taking a simple model as a paradigm.
Due to the noise the attractors become 'metastable' and the 
trajectory starts hopping between the different attractors. In the case of 
fractal basin boundaries, this hopping process consists of two steps.
In the first one the trajectory leaves the open neighborhood about
the attractor according to Arrhenius law \cite{grassberger:1989}, 
in the second one, the trajectory bounces around on the chaotic saddle 
before cascading again into the open neighborhood of an attractor. The 
general mechanism for the first step of the hopping process is the same for 
bi- and multistable systems. In both cases we find Arrhenius law; quanitative 
differences are due to different relative sizes of basins of attraction 
\cite{Kraut:1999}
However, for the attractor-hopping dynamics, our study points out a
major difference between bistable and multistable systems: While in
bistable systems the structure of the saddle separating the two
attractors (saddle point or chaotic saddle) does not play a role for the
hopping characteristics, it is essential for the attractor-hopping
in a system possessing a multitude of attractors.
The structure of the chaotic saddles influences the nature of the
hopping process, in particular it determines which transitions
between attractors are possible.
Bifurcations in the chaotic saddles lead to
changes in the 'accessibility' of attractors in the hopping process.
\\
Employing symbolic dynamics, we assign one symbol to each attractor,
thus transforming our hopping time series onto a symbolic string. As a
next step we compute both, the Shannon and the
topological entropy, which can be regarded as measures of complexity
for the hopping dynamics. As a system parameter is varied the
complexity of the hopping process changes beyond a certain threshold
value. We show, that this change can be related to a novel
bifurcation, namely a merging of two chaotic saddles accompanied by
the emergence of additional points filling the gap between the formely
separated saddles. This bifurcation is mediated by a snapback repellor
\cite{marotto:1978}, whose eigenvalues determine the scaling law for
the transient lifetimes on one chaotic saddle close to the bifurcation
point.
\\
Our basic prototype model of multistablity, which captures
the main features of highly multistable systems,
namely more than two final states (attractors) and a complexly
interwoven fractal basin boundary separating these states,
is given by the 10-fold iterate of two coupled logistic maps:
\begin{eqnarray} \label{coupled_log_map}
x_{k+1} & = & 1.0 - \alpha  x_k^{2} + \gamma ( y_k -x_k) \nonumber \\
y_{k+1} & = & 1.0 - \alpha  y_k^{2} + \gamma ( x_k -y_k). 
\end{eqnarray}
We fix $\gamma$, the coupling strength, at $\gamma = 0.29$, and vary
$\alpha$, the nonlinearity, in the range $ \alpha \in [0.72,0.755]$.
For this parameter set, there exists a stable period $10$-orbit
for the two coupled logistic maps themselves. Since we consider the 10-fold
iterate, our map exhibits 10 coexisting fixed point attractors, where 
each 5 lie above and below the symmetry axis $x=y$, respectively. Using 
only initial conditions below the symmetry axis, 
we can restrict our study to a system possessing 5 coexisting fixed point 
attractors, 5 repellors with two unstable directions and 10 saddle points.
Each of the 5 attractors is surrounded by an open neighborhood of
different size, in which all initial conditions converge to the
corresponding attractor inside this neighborhood. The basins of
attraction as a whole have a complex fractal structure due to a
homoclinic bifurcation which occurs already for smaller values of the
nonlinearity $\alpha$. As a next step we apply noise to the system,
using Gaussian, white noise with standarddeviation $\sigma$  added to
the $x$ and $y$ component of the 10-fold iterate of (\ref{coupled_log_map}). 
Small noise causes the system now to alternate between the different
states, where long periods close to a fixed point, comparable to
laminar motion, are interrupted by short, sudden bursts, which are
reminiscent to intermittent behavior.
During the bursts the motion takes place on the chaotic saddles
separating the attractors, until the trajectory is again being
injected in the neighborhood of one of the $5$ fixed points.
This complex dynamics is depicted in Fig.\ \ref{timeseries}.
According to our aim we are now interested in the characteristic
properties of the hopping dynamics.
Thus we address the question, how the complexity of the hopping dynamics
changes with the nonlinearity parameter $\alpha$, namely whether every 
fixed point has a positive transition probability to every other fixed
point, via a transient on a chaotic saddle. For this purpose, we employ 
the concept of symbolic dynamics. We use an encoding scheme in which
we assign one symbol to each attractor \cite{poon:1995}: Neglecting the
number of iterations the trajectory spends close to a fixed point, a
symbol is given for every fixed point and only after a jump out of an
attractor a new symbol is bestowed, according to the attractor where
the trajectory lands at. Using this scheme, we focus 
only on the structural properties of the jumps, not taking fully into
account the complete temporal evolution in each iteration.
As a quantitative measure of the complexity of the
symbol string we use the Shannon entropy, in analogy to the
Kolmogorov-Sinai entropy \cite{kolmogorov:1958}, given by
\begin{eqnarray} \label{ks-entropy}
h_{S} & = &\lim_{n \to \infty} \frac{H_n}{n} = \lim_{n \to \infty} ( H_{n+1} - H_{n}) \nonumber \\
  & = & \lim_{n \to \infty} \frac{1}{n} \left( - \sum_{ |S| = n} p(S) \log p(S) \right),
\end{eqnarray}
where $S = s_{1} s_{2} ... s_{n}$ denotes a finite symbol
sequence consisting of $n$ elements $s_{i}=1,2,...,5$, $p(S)$ its probability
of occurrence and $H_{n}$ the block entropy of block length $n$.
Numerically, the quantity $\lim_{n\to \infty} (H_{n+1}- H_{n})$
converges already for $n=1$, indicating that the hopping dynamics is a Markov
process of first order. This corresponds to the intuitive expectation, that
the dynamics possesses only memory of the last state it was dwelling on.
Higher correlations are suppressed by the long laminar-like motion.
Consequently a transition matrix between the different fixed points
can be constructed, where the entries are the probabilities for a transition
between two states.
\newline
Another interesting quantity to investigate is the topological  entropy.
One way of computing it is to built the Stefan transition matrix
\cite{derrida:1978}, where the entries of the $5 \times 5$ matrix
are $1$ and $0$, depending
on whether a transition took place or not, respectively. The logarithm of
the largest eigenvalue of this matrix yields the topological entropy.
In our numerical implementation, we use a certain, very small cutoff limit
of $0.001$ for the transition probabilities, below which we regard the
value as zero. Thus we neglect transition, having extremely 
small probabilities compared to the other transitions. 
However, the exact value of this cutoff does
not change the results significantly, as long as it is small enough. 
The entropies shown in Fig. \ref{entropy} remain unchanged for cutoff 
limits between $10^{-3}$ to $10^{-6}$.
It is important to note that, although the {\it escape times} out of
the stable states depend on the noise level, the
{\it transition probabilities} remain constant for a
wide range of noise values. Thus the considered entropies are not
affected by the noise amplitude for a rather large interval of noise
levels.
In Fig.\ \ref{entropy} we present the evolution of these two quantities
as the nonlinearity $\alpha$ changes. The curve for the topological
entropy appears to be a monotonic curve with a devil's staircase like
behavior \cite{biham:1989}.
Every time it increases, at least one new transition between hitherto
unconnected fixed points is created. Below the
homoclinic bifurcation $\alpha < 0.72$, where the basin boundaries are smooth,
only one transition per fixed point is possible, as the fixed points are
located on a closed curve made up by the unstable manifolds of the
saddlepoints. These unstable manifolds
impose a direction. Consequently, the topological entropy is zero.
Above a value of  $\alpha \approx 0.75$, all transitions can occur and
the maximum of the topological entropy is reached. The Shannon entropy
is always less than the topological entropy. Besides the overall increase
of both entropies, the plateau between $0.74 < \alpha < 0.744 $ suggests,
that there is a  change in the dynamical behavior beyond it.
This change is due to a change in the topological structure of the
system and relates to a new bifurcation as we show next.
\\
We compute the chaotic saddle, responsible for the chaotic dynamics,
by the PIM-triple (proper interior maximum) method \cite{nusse:1989}.
The result is depicted in Fig.\  \ref{saddle1},  and
\ref{saddle2}, where the chaotic saddles are
shown for two different values of $\alpha$ together with the
$4$-fold preimages $(+)$ of one repellor $R$ marked with a $\Box$.
Slightly above the homoclinic bifurcation ($\alpha = 0.725$)
the saddles are very thin, consisting of two separate rings.
At a value of $\alpha = 0.755$, there exists only one large piece
of the nonattracting chaotic set (not shown).
This results from the saddle merging, which took place at
$\alpha \approx 0.7430$  (Fig.\ \ref{saddle2}). The two bands
of chaotic saddles merge and the repellor, formerly located ouside
of the saddles, is now embedded in the large saddle, including
some of its preimages.
The merging  is accompanied by the successive filling of the
formerly empty gap between the two saddles. A similar effect has been
observed in  \cite{chin:1992} where it is called 'spilling', in 
chaotic scattering \cite{lai:1993}, and in the
gap filling crisis \cite{szabo:1996}. In contrast to the
latter this happens smoothly in the saddle merging.

\begin{figure}[h]
\begin{center}
\epsfig{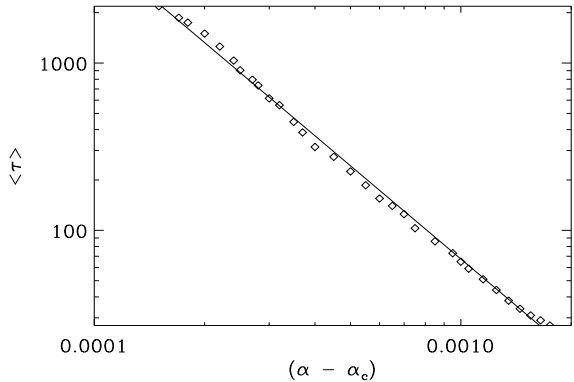}
\vspace{-0.2cm}
\caption{Scaling of the transient lifetime to stay in one ring of the
chaotic saddle with the nonlinearity $\alpha$ and $\alpha_c = 0.7428$.
The slope of the log-log plot is $\epsilon = 1.86$.}
\label{scaling_escape}
\end{center}
\end{figure}

However, the actual bifurcation takes place at a slightly lower value  of
$\alpha \approx 0.7428$ (not shown).
At this parameter value some of the preimages of the marked repellor touch
its unstable manifold, which constitutes the border of the chaotic saddle.
Through this mechanism, a snapback repellor has developed.
A fixed point p is called a snapback repellor, if
i) all eigenvalues of p have absolute values larger than 1, ii) there exists
a $q \in W_{loc}^{u}(p)$, the unstable manifold of p, such that
the M-fold iterate of the map ${\cal F}^M(q) = p$ for some positive
integer M and iii) $\det D {\cal F}^M(q) \neq 0 $\cite{marotto:1978}.
While conditions i) and iii) are true for all $\alpha$ considered here,
it is condition ii) that becomes fulfilled during the bifurcation,
as the unstable manifold of the repellor touches its preimages.
This bifurcation is responsible for the emergence
of the additional points in the gaps, (see Fig.\ \ref{saddle2}).
At a slightly higher value of  $\alpha \approx 0.7430$ the preimages
touch not only the unstable manifold, constituting the border of the
chaotic saddle, but the saddle itself.
As the chaotic saddle is an invariant set, the repellor becomes thus
embedded in the saddle and the merging of the two rings has taken place.
This merging has an important consequence for the dynamics of the
attractor-hopping. Not only jumps between neighbors, but arbitrary
jumps become, in principle,  possible.
Above the critical value, the connected pieces of the chaotic saddle
enable the trajectory to jump between the formerly separated parts,
thereby causing a sharp increase in the number of the possible transitions,
as can be seen from the topological entropy above $\alpha = 0.7430$ in 
Fig.\ \ref{entropy}.
\\
Since the bifurcation described above can be
considered as a kind of crisis, it is interesting to see, whether
the scaling laws for the characteristic transient lifetimes
\cite{ott:1993} apply in this case as well. We compute the average
escape times out of one (still separated) chaotic saddle ring slightly
below the bifurcation. In Fig.\ \ref{scaling_escape} the data are plotted,
resulting in a scaling relation:
\begin{equation}
\langle \tau \rangle \sim (\alpha - \alpha_c)^{-\epsilon},
\end{equation}
where $\alpha_c = 0.7428$. The best fit to the data yields an
exponent $\epsilon = 1.86 $. That value can be reproduced by
\begin{eqnarray}
\epsilon & = & 1/2 + ( \ln(\beta_1) / \ln(\beta_2)) \\
\epsilon & = & 1/2 + (\ln (2.45) / \ln(1.9) ) = 1.85,
\end{eqnarray}
where  $\beta_1$ and $\beta_2$ are the expanding eigenvalues of the
mediating repellor.
This coincidence has been confirmed also for other values of the
coupling strength $\gamma = 0.285,0.287,0.292$. It is worth mentioning that
in the original derivation of this formula, which applies to the
heteroclinic tangency crisis, $\beta_1$ and $\beta_2$ are the expanding
and contracting eigenvalues of the mediating saddle point \cite{grebogi:1986}.
However, in our case both, $\beta_1$ and $\beta_2$, are  expanding 
eigenvalues of the mediating repellor. We conjecture this relation to
hold for the general case of a crisis caused by a snapback repellor.
The reason, that a law for the heteroclinic tangency is supposed to
hold in a homoclinic case of a snapback repellor is, that the
mechanism for approaching the fixed point is different here. The fixed point
is not reached by successive iterations, but suddenly with a jump of
noninfinitesimal size from a finite distance.
\\
In conclusion, we have studied a simple paradigmatic map possessing a
multitude of coexisting stable states under the influence of small noise.
We have shown that the nature of the attractor-hopping process depends,
in contrast to the bistable case, crucially on the structure of the
chaotic saddles separating the attractors.
Bifurcations of the saddles lead to a change in the hopping dynamics
which is manifested in a sudden increase of dynamical entropies.
In particular we described a novel bifurcation, a merging of two
chaotic saddles involving a snapback repellor, which results in a change
of the 'accessibility' of the attractors.
Finally, a conjecture for the scaling law of the transient lifetimes on the
chaotic saddle has been provided.
The hopping dynamics, the bifurcation and the scaling law should
be observable in experimental realizations of such systems.
We expect this bifurcation  also to occur in systems with three or
more dimensions, where the mediating saddle has at least two unstable
directions.
\\
We acknowledge Y. Maistrenko for valuable discussions and the DFG
for financial support.


\begin{thebibliography}{10}


\bibitem{Wiesenfeld:1990}

K.~Wiesenfeld, C.~Bracikowski, G.~James, and R.~Roy, Phys. Rev. Lett. {\bf 65},
  1749 (1990); M.~Brambilla, L.~A. Lugiato, V.~Penna, F.~Prati,
  C.~Tamm, and C.O. Weiss, Phys. Rev. A {\bf 43}, 5114 (1991).

\bibitem{Marmillot:1991}
P.~Marmillot, M~Kaufmann, and J.-F. Hervagault, J. Chem. Phys. {\bf 95}, 1206
  (1991); J.~Liu and S.~K. Scott, Dyn. \& Stability Syst. {\bf 8}, 273 (1993).

\bibitem{schiff:1994}
S.~Schiff, K.~Lerger, D.~H. Duong, T.~Chang, M.~L. Spano, and W.~L. Ditto,
  Nature {\bf 370}, 615 (1994); J.~Foss, A.~Longtin, B.~Mensour, and J.~Milton, Phys. Rev. Lett. {\bf 76}, 708 (1996).

\bibitem{Prengel:1994}
F.~Prengel, A.~Wacker, and E.~Sch\"oll, Phys. Rev. B {\bf 50}, 1705 (1994).

\bibitem{Chern:1991}
C.~S. Chern and Lin~I, Phys. Rev. A {\bf 43}, 1994 (1991).

\bibitem{kim:1997b}
S.~Kim, S.H. Park, and C.S.Ryu, Phys. Rev. Lett. {\bf 79}, 2911 (1997).

\bibitem{Kaneko:1997}
K.~Kaneko, Phys. Rev. Lett. {\bf 78}, 2736 (1997).

\bibitem{Kraut:1999}
S.~Kraut, U.~Feudel, and C.~Grebogi, Phys. Rev. E {\bf 59}, 5253 (1999).

\bibitem{hondou:1995}
T.~Hondou and Y.~Sawada, Phys. Rev. Lett. {\bf 75}, 3269 (1995).

\bibitem{kim:1997a}
S.~Kim, S.H. Park, and C.S.Ryu, Phys. Rev. Lett. {\bf 78}, 1616 (1997).

\bibitem{ikeda:1989}
K.Ikeda, K.~Matsumoto, and K.~Otsuka, Prog. Theor. Phys. Suppl. {\bf 99}, 295
  (1989); K.~Wiesenfeld and P.~Hadley, Phys. Rev. Lett. {\bf 62}, 1335 (1989).

\bibitem{Arecchi:1990}
F.~T. Arecchi, G.~Giacomelli, P.~L. Ramazza, and S.~Residori, Phys. Rev. Lett.
  {\bf 65}, 2531 (1990).

\bibitem{grassberger:1989}
P.~Grassberger, J. Phys. A {\bf 22}, 3283 (1989).

\bibitem{marotto:1978}
F.~R. Marotto, J. Math. Anal. Appl. {\bf 63}, 199 (1978).

\bibitem{poon:1995}
L.~Poon and C.~Grebogi, Phys. Rev. Lett. {\bf 75}, 4023 (1995).

\bibitem{kolmogorov:1958}
A.~N. Kolmogorov, Dok. Acad. Nauk SSSR {\bf 119}, 861 (1958); Ya.~G. Sinai, Dok. Acad. Nauk SSSR {\bf 124}, 764 (1959).

\bibitem{derrida:1978}
B.~Derrida, A.~Gervois, and Y.~Pomeau, Ann. Inst. Poincar\'e A {\bf 29}, 305
  (1978); B.-L. Hao, {\it Elementary Symbolic Dynamics}, World Scientific,
  Singapore, 1989.

\bibitem{biham:1989}
O.~Biham and W.~Wenzel, Phys. Rev. Lett. {\bf 63}, 819 (1989);
W.~Breymann and J.~Vollmer, Z. Phys. B {\bf 103}, 539 (1997);
Y.-C.~Lai and K.~{\.Z}yczkowski and C.~Grebogi, Phys. Rev. E {\bf 59}, 5261 (1999).

\bibitem{nusse:1989}
H.~E. Nusse and J.~A. Yorke, Physica D {\bf 36}, 137 (1989).

\bibitem{chin:1992}
W.~Chin, I.~Kan, and C.~Grebogi, Rand. \& Comp. Dynamics {\bf 1(3)}, 349
  (1992).

\bibitem{lai:1993}
Y.-C. Lai, C.~Grebogi, R. Bl\"umel and I. Kan, Phys. Rev. Lett. {\bf 71},
  2212 (1993).
  
\bibitem{szabo:1996}
K.~G. Szab{\'o}, Y.-C. Lai, T.~T{\'e}l, and C.~Grebogi, Phys. Rev. Lett. {\bf
  77}, 3102 (1996); K.~G. Szab{\'o}, Y.-C. Lai, T.~T{\'e}l, and C.~Grebogi,
  Phys. Rev. E {\bf 61}, 5019 (2000).

\bibitem{ott:1993}
E.~Ott, {\it Chaos in dynamical systems}, Cambridge University Press,
  Cambridge, 1993.

\bibitem{grebogi:1986}
C.~Grebogi, E.~Ott, and J.A. Yorke, Phys. Rev. Lett. {\bf 57}, 1284 (1986).

\end{thebibliography}
\end{document}